 \documentstyle[12pt,a4]{article}

 \author{ Stefano Battiston $^{(1)}$ G{\'e}rard Weisbuch $^{(1)}$
     and Eric Bonabeau $^{(2)}$ \\ 
   {\small (1) Laboratoire de Physique Statistique, ENS, 24 rue Lhomond,
            75005 Paris, France}\\
   {\small (2) Icosystem Corp., 545 Concord Av. Cambridge, MA 02138, USA}\\[0.1 cm]
   {\small Correspondence to: stefano.battiston@ens.fr, tel +33144323623, fax +33144323433 }\\[0.1 cm]
 }

 \title{Decision spread in the corporate board network}
 
\begin{document}

 \input epsf
 \maketitle

\abstract {Boards of large corporations sharing some of their directors
are connected in complex networks. Boards are responsible for
corporations' long-term strategy and are often involved in decisions
about a common topic related to the belief in economical growth or
recession.

We are interested in understanding under which conditions a large
majority of boards making a same decision can emerge in the network. We
present a model where board directors are engaged in a decision making
dynamics based on "herd behavior". Boards influence each other through
shared directors.

We find that imitation of colleagues and opinion bias due to the
interlock do not trigger an avalanche of identical decisions over the
board network, whereas the information about interlocked boards'
decisions does. There is no need to invoke global public information, nor
external driving forces.

This model provides a simple endogenous mechanism to explain the
fact that boards of the largest corporations of a country can, in
the span of a few months, take the same decisions about general
topics.} \newline

PACS: 89.75 -k; 89.65 -s

\textit{Keywords}: social networks, opinion dynamics, directorate
interlock, Ising model.\newline

\section{ Introduction}
We present a model for the diffusion of decisions in a network of
corporate boards linked by shared board directors. Boards of large
corporations share common directors with each other forming complex
networks, often characterized by Small World properties \cite{Davis02}.

\begin{figure}[tbh]
    \centerline{ \epsfxsize=130mm\epsfbox{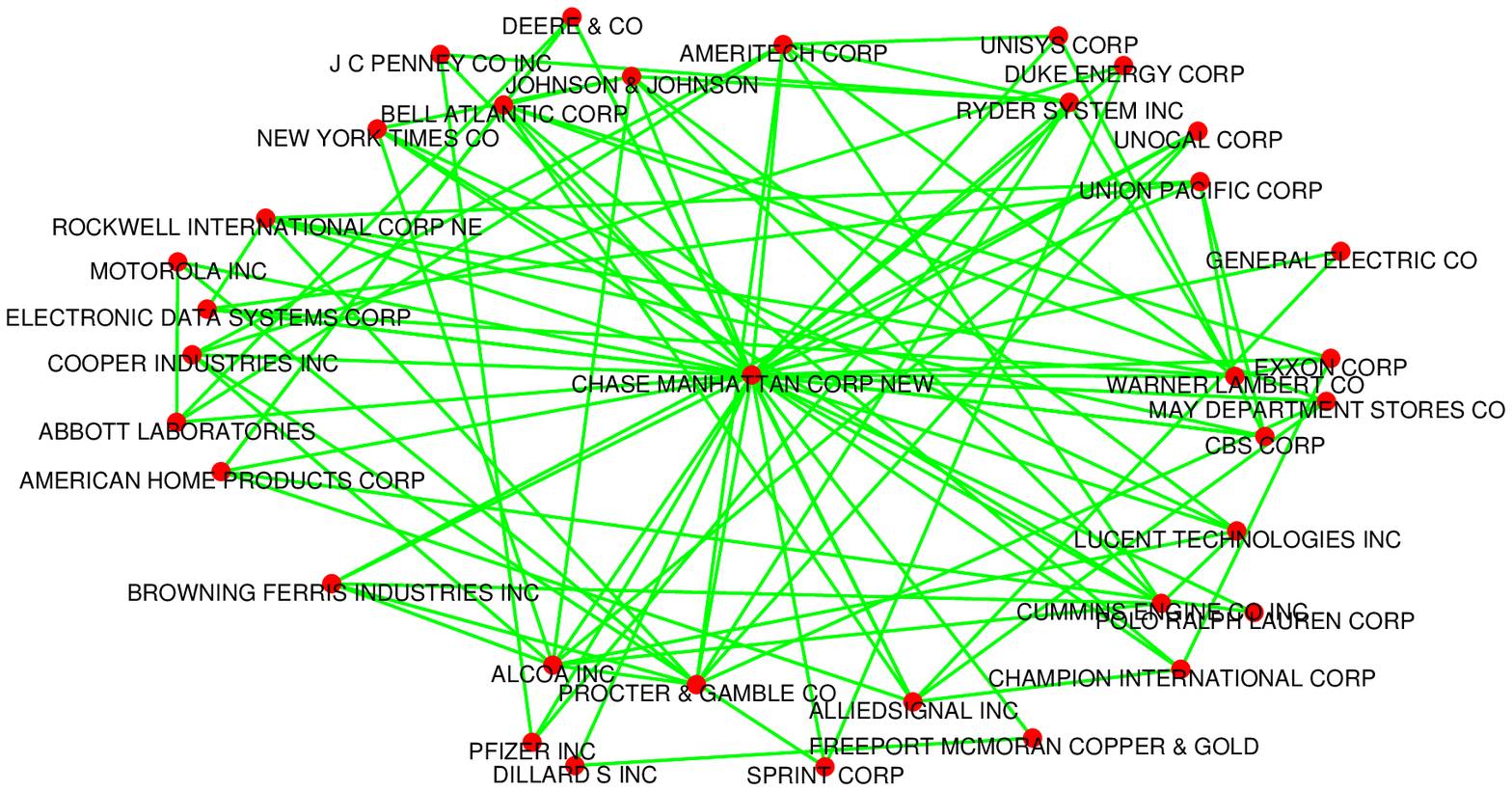}}
    \centerline{\parbox{320pt}{ \caption{
The network of boards at 1 degree of separation from Chase Manhattan
Bank's board (CMB net 1).
     }\label{ChaseNet1Labels}
     }}
\end{figure}

An example of board network is shown in figure \ref{ChaseNet1Labels}:
nodes represent boards of directors, two boards are connected by an edge
if there is at least one director sitting on both boards at the same
time. The network represents the boards at 1 degree of separation (in
term of edges) away from the board of Chase Manhattan Bank. For instance
the boards of Abbott Laboratories and Chase Manhattan Bank are connected
because they share one director.

We are interested in the influence of such networks on the decisions made
by boards. The role of boards is to make decisions about the long-term
strategy of the corporations. There are essentially two kinds of
decisions a board is faced to:
\begin{itemize}
\item

Decisions regarding topics specific to the board, such as the appointment
of a vice president, for which we can assume that different boards don't
influence each other. We previously studied \cite{Battiston 2003} the
role of subsets of well connected directors on decisions of this type.

\item
By contrast, there are decisions about
 topics of general interest to the economy such as whether to increase or
decrease investments in development or in advertisement, which
 depend on the belief in
economical growth or recession. Decisions of equivalent generality
concern the adoption of governance practices. In these cases, decisions
previously made in some boards might influence other boards, through the
presence of shared directors. The present paper is devoted to the
dynamics of this kind of decisions across the board network.
\end{itemize}

The second issue has already been studied in sociology. Haunschild in an
article \cite{Haunschild} about the impact of interlock on US corporate
acquisition activity in the 80', demonstrates the role of
inter-organizational imitation of managers. The article by Davis and
Greve \cite{Davis96} concerns the diffusion of corporate governance
practices such as the "poison pill" among the US largest corporations in
the 80' (The poison pill is a counter measure against hostile takover
allowing ``target shareholders to acquire shares at a 50 perc. discount
if an acquirer passes a certain ownership threshold'').

Davis and Greve analyse a large set of empirical data and show that the
poison pill was primarily mediated by director interlock, with a
spreading time scale of the order of one year. In order to explain the
observed spreading dynamics, the authors develop a contagion model in
which boards are agents with a certain probability to imitate an earlier
adopter of the practice. The influence of a board on an interlocked board
is reduced or enhanced by a number of factors, discussed in detail by the
authors, such as similar industrial sector, firm size, firm performance,
concentration of ownership of both companies involved.

We here present a model where agents are instead board directors engaged
in a decision making dynamics. Inter-boards influence takes place through
shared directors. We are not taking into account the heterogeneity of
inter-boards influences. We are interested in understanding under which
conditions a large majority of boards making a same decision can emerge.
The decision making process is based on the hypothesis that agents tend
to follow the majority of the agents to whom they are connected.

This kind of behavior is not perfectly rational in the sense of
economics and is known as "herd behavior" \cite{Orlean}
\cite{Follmer}, but presents formal analogies with the Ising model
often used for social system models, \cite{Galam 82} \cite{Galam
91} \cite{Weisbuch 99}.

Models \cite{Galam 82} \cite{Galam 91} \cite{Weisbuch 99} are
based on binary opinion dynamics. The basic updating process is
the same as described here by equations \ref{eq:mod1:1} and
\ref{eq:mod1:2}, but interaction topologies are simplified to
either full connectivity or lattice connectivity.

Unlike the cited works, in the present model the dynamics takes
place on an empirical etherogeneous network organized in
interconnected groups. This is the director network of Fortune
1000 corporations (data are kindly provided by Davis
\cite{Davis02}). The connectivity within a group (a board of
directors) and between groups is highly etherogeneous. Moreover,
the dynamics takes places in different groups at different times.

Furthermore, we explore two different scenarios for the influence
exerted by groups upon each other. In the first scenario the
information about other groups' decisions only affects the initial
opinions of interlocking directors. In the second scenario such
information is taken into account by all directors during the
whole board meeting.

Our main result in this paper is that the way this information is
taken into account by directors at a local level, determines
whether, at a global level, it can emerge a large majority of
boards making the same decision.

\subsection{ Building a model for the spread of decisions }
We here focus on a time scale of a few months, just the time for boards
to hold a couple of meetings. Typically we have in mind fast decisions
concerning investment and advertisement strategies for the next six
months. Decisions of this type are relevant for the macro-economics of a
country, especially if corporations tend to make similar decisions in
connection to external dramatic events such as crises or wars. It is
therefore important to understand what structural parameters of the
network determine the number of boards adopting a given decision. Of
course this is a formidable task to accomplish, so we start considering
in this paper a simple situation in which decisions are binary and the
most central board of the network takes a decision in the first place and
the other boards meet afterwards. Because of the interlock, the decision
made in a given board is influenced by the presence of some directors
that were present at a previous meeting on another board.

Let us assume that each board has to make a binary decision. The opinion
of director $i$ is represented by a binary variable $s_{i}=\pm1$. For the
decision making dynamics in each board meeting, we assume a survey-like
mechanism, as in \cite{Battiston 2003}:  each director updates his
opinion based on the average opinion of the other directors, according to
a logit probability function (see below).

Boards have meetings once per month, according to a given schedule.
Directors start from an initial opinion $s_{i}=\pm 1$ at the beginning of
the meeting. By the end of a meeting, directors adopt the opinion that
agrees with the decision made by the board.

Therefore, directors serving on several boards start a new meeting with
an initial opinion adopted at the last meeting in another board. In this
way the decision made previously in a board can influence the decision
making dynamics in another board.

After first studying the decision spread on some simple test-bed networks
we concentrated on the real network of the boards 1 and 2 degrees of
separation away from the board of Chase Manhattan Bank in 1999.

The most central board of the network meets first and the other boards
meet afterwards, according to a given schedule.

The topological structure of the network is known from the names of the
directors in the boards of the set of companies under consideration. On
the other hand, we do not know the magnitude of the influence of a board
on another one.

We here assume that the influence of one board on a connected board
varies in accordance with the number of directors shared by the two
boards (the interlock). For sake of simplicity we don't take into account
those specific factors such as: some boards are very influential over
other ones because of differences in prestige, revenue, economic
performance, expertise in a specific field, see \cite{Davis96} for a
detailed analysis.

We here present a first model in which the influence of a board $b_{1}$
on another board $b_{2}$ is due only to the fact that those directors of
$b_{1}$ which sit also in $b_{2}$, will support, at the beginning of the
meeting in $b_{2}$, the choice adopted by $b_{1}$. We study how the
pattern of decisions depends on the value of the average interlock.

In the second model the influence of the interlock is increased
by two factors.
\begin{itemize}
\item
  Inside the board, directors sitting also on common outside boards have
a larger mutual influence than other directors.
\item
  Decisions made by other connected boards can be taken into
account as external signals by all board members.
\end{itemize}

\section{ Model 1}

\subsection{Inside board dynamics}

We here describe the simplest model. Directors meet, discuss and vote.

In any new meeting directors' opinions are initialized as follows:
\begin{itemize}
\item Directors participating at a meeting for the first time
are initialized with opinions randomly chosen among 0 or 1;
\item all other
directors start with the opinion they acquired at the previous meeting they
participated at,
\item except for directors from the most central node $b_{c}$
which start at any new meeting with the opinion that won in $b_{c}$.
\end{itemize}

At each time step, a director randomly selected is informed of the
opinions of all other directors in the board at time t, and average them
to evaluate a field $h_{i}$:
\begin{equation}\label{eq:mod1:1}
    h_{i}=\frac{1}{n}\sum_{j=1}^{n}s_{j}
\end{equation}
$n$ being the size of the board. He updates his opinion according to
a probabilistic choice function of the field.
 The probability that director $i$ takes
some opinion $\pm1$ at time $t+1$ is given by:
\begin{equation}\label{eq:mod1:2}
    P\{s_{i}(t+1)=\pm1\}=\frac{ \exp(\pm \beta h_{i}(t))}
                             {\exp(\beta h_{i}(t))+\exp(-\beta h_{i}(t))}
\end{equation}
 The updated opinion is expressed by the director and is taken into
account by later sampled directors.

The average opinion $m$ of the board, is a function of time during each
meeting and eventually converges at a value $m^{*}=\pm1$ for $\beta\gg1$.
In all simulations we set $\beta=4$ such that the board converges to
unanimity in less than 30 steps, i.e. in roughly three average sampling
of each director.
 We take the final value of $m^{*}$ as the decision of the
board. We will write $m$ for $m^{*}$ in the following for sake of
simplicity.

\subsection{ Boards' network dynamics }
  The central node meets first and the decision made is by
  convention $+1$.

Boards are assumed to meet monthly and to discuss again the same topic.
Each step represents a month, and hence a cycle of board meetings. The
central board meet first, followed by the nodes on each surrounding
layer, in order of distance from the center. Unless specified, the
meeting schedule described here will be assumed in the following for
model 1 and for model 2.

Within the model described so far, the whole network converges to a
stable value of the average decision $M^{*}$ in 2-3 steps, which is
reasonable. In fact it is conceivable that some actions require more than
one decisional step, to be fully approved.

If a board happens to make a decision -1, it is quite reluctant to change
in the next. This model differs from usual contagion models in that the
risk of infection or adoption of an innovation may be diminished by an
internal dynamics of the board.

We will write $M$ for $M^{*}$ in the following for sake of simplicity. We
run several simulations with varying random seeds per set of parameters
and
 we monitor the following variables:
\begin{itemize}
  \item The average decision $<m_{k}>$ of the board $k$.
  Because $m_{k}=\pm1$ at each run, the probability $p^{+}_{k}$ that $m_{k}=+1$
  is given by: $p^{+}_{k}=\frac{1+<m_{k}>}{2} $
  \item The average decision $M$ of the whole board network over several runs.
  We will call $P^{+}$ the fraction of boards that make decision +1,
  averaged over several runs.
\end{itemize}

\section{ Results}
\begin{figure}[tbh]
    \centerline{ \epsfxsize=70mm\epsfbox{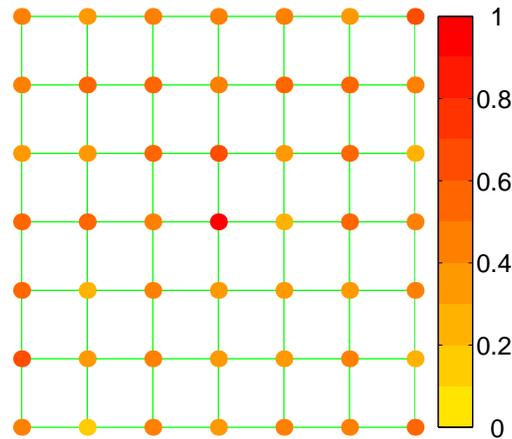}}
    \centerline{\parbox{320pt}{ \caption{
Square lattice of 7X7 nodes. A node represents a board of n=10 directors,
a link represents the existence of some shared directors between two
boards. Nodes color corresponds to the value of the probability $P^{+}$
of adopting decision +1, when the fraction $\gamma$ of shared directors
between two connected boards is 0.4.
     }\label{figLatt_g04}
     }}
\end{figure}

 We present the results of model 1 on different networks of
 increasing connectivity.
First, with consider a square lattice (\textbf{figure \ref{figLatt_g04}})
with an odd number of vertices on the edge, so that there is a node in
the center of the lattice. A node represents a board of n=10 directors, a
link represents the existence of some shared directors between two
boards. The fraction of shared directors between two connected boards is
a parameter $\gamma$ that we can control. For each pair of connected
boards we can impose that they share a larger fraction of directors by
replacing a director of the first board with a director of the second one
(provided that each director sits only on one chair!). It is very
important to remark that increasing the number of shared directors per
pair of boards, introduces at some point some links between boards that
were not connected before. For instance at the end when each board is
sharing all its directors with its four initial neighbouring boards, in
fact all boards must share the same 10 directors and therefore the graph
must be complete. With no interlock, all nodes are isolated and
$\gamma=0$.\newline

\begin{figure}[tbh]
    \centerline{ \epsfxsize=80mm\epsfbox{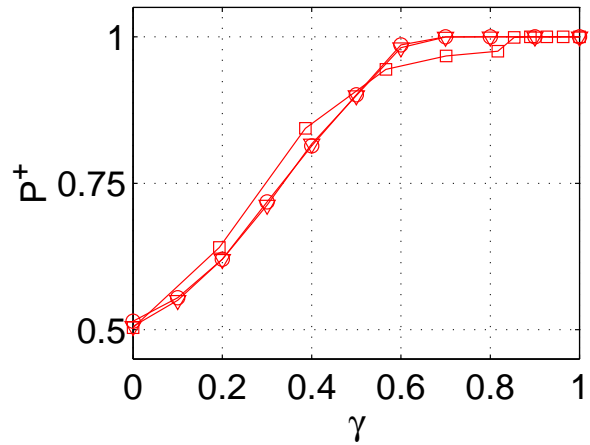}}
    \centerline{\parbox{320pt}{ \caption{
Probability of adoption of decision +1 as a function of the fraction
$\gamma$ of shared directors between two connected boards. circles:
lattice; triangles: star network; squares: the network CMB net 1 shown
in figure \ref{ChaseNet1Labels}.
     }\label{figPvsG_toys}
     }}
\end{figure}

The fraction $P^{+}$ of boards that adopted decision +1 is plotted as a
function of $\gamma$ in \textbf{figure
\ref{figPvsG_toys}}(circles).\newline

\begin{figure}[tbh]
    \centerline{ \epsfxsize=80mm\epsfbox{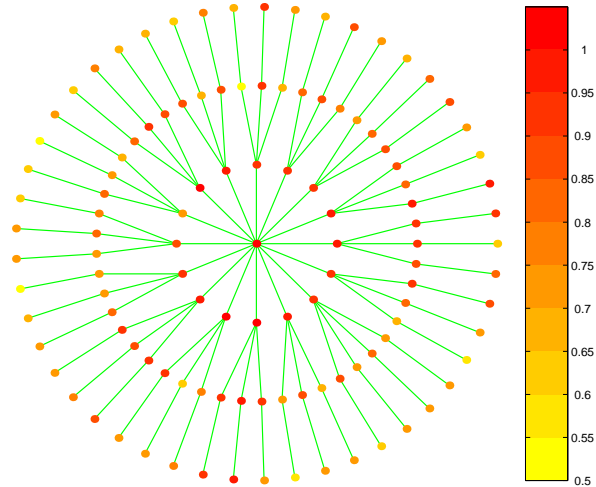}}
    \centerline{\parbox{320pt}{ \caption{
Star network of 115 nodes. The central node has connectivity degree 16.
The nodes in layers 2,3,4 have connectivity degree 4,2,1 respectively.
Nodes color corresponds to the value of the probability $P^{+}$ of
adopting decision +1, when the fraction $\gamma$ of shared directors
between two connected boards is 0.4.
     }\label{figStar_g10}
     }}
\end{figure}

The same procedure has been applied to a star network (\textbf{figure
\ref{figStar_g10}}) and to the network of boards 1 degree of separation
away from the CMB's board (\textbf{figure \ref{ChaseNet1Labels}}). Plots
of $P^{+}$ are shown in \textbf{figure \ref{figPvsG_toys}}(triangles and
squares, respectively).

$P^{+}$ increases linearly with the average fraction $\gamma$ of
directors shared by any two boards until it saturates for $\gamma$ around
0.6. The value of $\gamma$ observed in real board networks is $\gamma=.1\pm.02$,
which corresponds to $P^{+}\simeq 0.6$, i.e. 60 per cent chances to adopt
decision +1, compared to 50 per cent chances in absence of interlock
($P^{+}=0.5$ for $\gamma=0$).

As a first result, within the hypothesis of this model the degree of
interlock observed in the real CMB's network is low compared to the one
required to a have a large majority of boards adopting a same decision.
Results will be further analysed in the discussion section.

\section{Model 2: the influence of lobbies and information }
 The second model shares the same general dynamics as the first model
 in terms of initial conditions and sequences of updating directors
 and boards opinion, but the field variables $h_i$ are increased by
 two terms taking into account mutual influence of well-connected directors
 and the propagation of information from boards which met previously.

The central question about the role of the interlock is how it influences
the decision making dynamics (\cite{Carpenter}, \cite{Fich}). In a
previous paper (\cite{Battiston 2003}) we proposed a mechanism that we
re-introduce in the present model. We assume that two directors in the
board who also serve
 in another outside board are likely to take each other's opinion into account
  more seriously than the opinion of directors with whom they don't have
  additional professional relationships. These directors form a graph
  called \textit{interlock graph of the board}, for simplicity called
  "lobby" in the following.
 In the equation for the field felt by directors who participate in a lobby
  we introduce a coupling factor.
  Director $i$ feels a field $h_{i}^{0}$ defined as follows:
\begin{equation}
    \label{eq:mod2:1}
    h_{i}^{0}=\frac{1}{n}\sum_{j=1}^{n}(1+\alpha_{1} J_{ij}) s_{j}
\end{equation}
where n is the size of the board, $J_{i,j}$ is the number of
\textit{outside} boards on which directors $i$ and $j$ sit together,
$\alpha_{1}$ is a parameter modulating the mutual influence of directors
in the lobby.\newline

Of course directors take into account their own opinion: by definition
$J(i,i)$ should be the total number of outside boards where director $i$
sits. This is much larger than the number of outside boards a director
can \textit{share} with another one. So in order to avoid giving a too
big weight to the opinion of a director himself compared to the opinion
of his colleagues, $J(i,i)$ is set as the number of outside boards where
director $i$ serves with at least one other director of the board. In
summary, a director with several appointments, takes into account his own
opinion more than that of his colleagues. Among his colleagues then, he
takes into account their opinions based on how many professional
relationships he holds with them.\newline

As a consequence of the dynamics, directors belonging to a connected
component of a lobby tend to have the same opinion. As the lobby is a
graph consisting of one or more connected components, it must be kept in
mind that different connected components may have different opinions and
their effects on the whole board could partly cancel out.\newline \par

Furthermore, information about what decision other boards have made so
far, can reach a board thanks to the interlocking directors. Therefore,
their influence is twofold: on one hand they have an initial opinion
which reflects the decision made in another board (model 1). They could
just support that same decision, keeping confidential the information
about the fact that the other board has made that decision. On the other
hand they can reveal this information to the other directors, which will
take it into account to form their opinion. We model this by means of a
second term in the equation for the field. This is a sort of external
field $h^{e}$ acting on all directors of board $k$, but specific to the
board $k$. It equals the sum of the decisions made so far in the boards
connected through interlock to the board $k$, weighted by the number of
shared directors:
\begin{equation}
    \label{eq:mod2:2}
    h^{e}=\sum_{l=1}^{deg(k)} J^{B}_{kl}b_{l}
\end{equation}
The sum is running on the boards interlocked to board $k$, deg(k) is the
connectivity degree of board $k$,
  $J^{B}_{kl}$ is the number of directors shared by boards $k$ and $l$,
   $b_{l}$ is the decision made by the board if it has already met,
   $b_{l}=0$ otherwise.\newline
Again we don't include any factor of prestige for the boards.
Nevertheless, a board is more influential on another if it has more
directors on it. It is important to stress the fact that boards that have
not yet met do not influence the decision.
\newline The information
reaching a board can then be different from the information about what
all other boards, interlocked or not, have decided so far.\newline

The total field acting on director $i$ is:
\begin{equation}
    \label{eq:mod2:3}
    h_{i}=h_{i}^{0}+h^{e}=\frac{1}{n}\sum_{j=1}^{n}(1+\alpha_{1} J_{ij}) s_{j}+
       \alpha_{2} \sum_{l=1}^{deg(k)} J^{B}_{kl}b_{l}
\end{equation}
where $\alpha_{2}$ is a parameter modulating the influence of the
information about other board's decisions.\newline

We are now interested in the behavior of the system in the space of the
parameters $\alpha_{1}$ and $\alpha_{2}$.

\section{ Results }
We have run model 2 on two real board networks: the boards that are 1
degree of separation away from Chase Manhattan Bank's (CMB) board, and
the boards 2 degrees of separation away from CMB's board. We refer to
them in the following as CMB Net 1 and  CMB Net 2, respectively. CMB Net
1 consists of 35 boards, with average size 12, and average fraction of
shared directors equals to 0.1. The average connectivity degree is 5. CMB
Net 2 consists of 277 boards, with average size 12, and average degree 9,
average fraction of shared directors 0.1.

We varied parameters $\alpha_{1}$ and $\alpha_{2}$ in order to understand
what can be the impact of lobbies and the impact of information on the
decision making process over the board network.

\begin{figure}[tbh]
    \centerline{ \epsfxsize=140mm\epsfbox{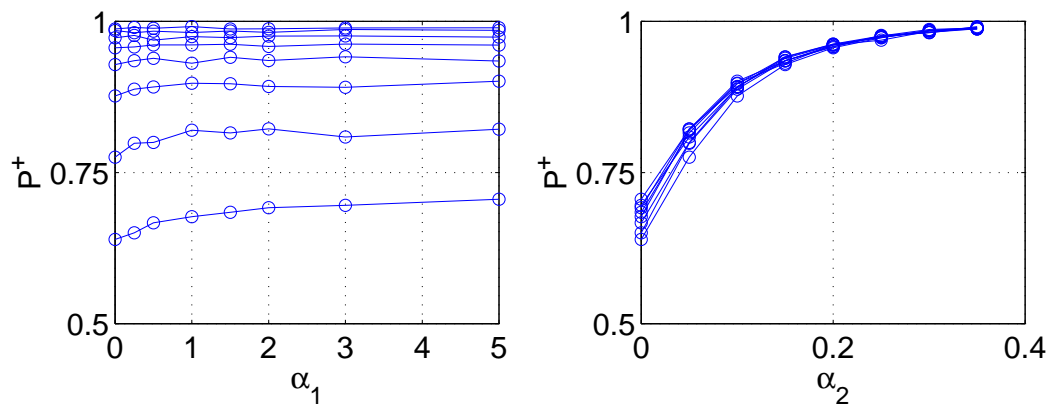}}
    \centerline{\parbox{320pt}{ \caption{
\textbf{left}: Probability $P^{+}$ of making decision +1 as a function
of the influence $\alpha_{1}$ of common appointments in outside boards.
Each curve corresponds to a fixed value of the influence $\alpha_{2}$
of the information about connected boards' decisions.
\textbf{right}: $P^{+}$ as a function of $\alpha_{2}$.
     }\label{figP_ChNet1}
     }}
\end{figure}

\begin{figure}[tbh]
    \centerline{ \epsfxsize=80mm\epsfbox{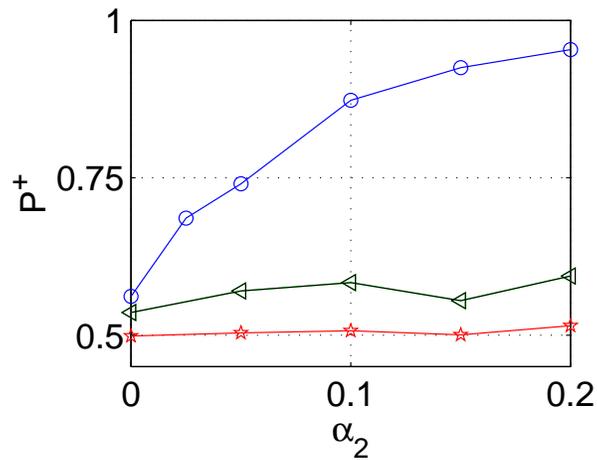}}
    \centerline{\parbox{320pt}{ \caption{
Probability $P^{+}$ of adopting decision +1 as a function of the influence
$\alpha_{2}$ of information about connected boards' decisions. Results on 100 runs.
Three meeting schedules.
\textbf{Circles}: first CMB, then boards at 1 degree of separation from
CMB, then those at 2 degree of separation . \textbf{Triangles}: CMB meets
first, then the other boards meet with a schedule randomly chosen at
every run. \textbf{Stars}: the schedule of all boards, including the
first, is randomly chosen at every run.
     }\label{figP_ChNet2}
     }}
\end{figure}

Results of simulations on CMB Net 1 and CMB Net 2 are given in
\textbf{figure \ref{figP_ChNet1}} and \textbf{figure \ref{figP_ChNet2}}
respectively.

For CMB Net 1, increasing the influence $\alpha_{1}$ of common
appointments in external boards from 0 to 1 while keeping $\alpha_{2}=0$
(\textbf{fig. \ref{figP_ChNet1} left}), leads to the increase of $P^{+}$
from $0.63\pm0.01$ to $0.68\pm0.01$. On the other hand $P^{+}$ increases
dramatically with the influence $\alpha_{2}$ of the information about
connected boards' decisions (\textbf{fig. \ref{figP_ChNet1} right}). For
high values of $\alpha_{2}$ there is no more dependence on $\alpha_{1}$.

When $\alpha_{1}=0$ and $\alpha_{2}=0$, then model 2 is formally
equivalent to model 1. We recall that we have run model 1 on a network
with the same topology as CBM Net1, but with an homogeneous fraction
$\gamma$ of shared directors. When $\gamma$ equals the average fraction
of shared directors observed in CBM Net1, then model 1 yields
$P^{+}=0.6$, close to $P^{+}=0.63$ found with model 2. So the
etherogeneity of the fraction of shared directors doesn't seem to play a
significant role on the network as a whole.

\begin{figure}[tbh]
    \centerline{ \epsfxsize=80mm\epsfbox{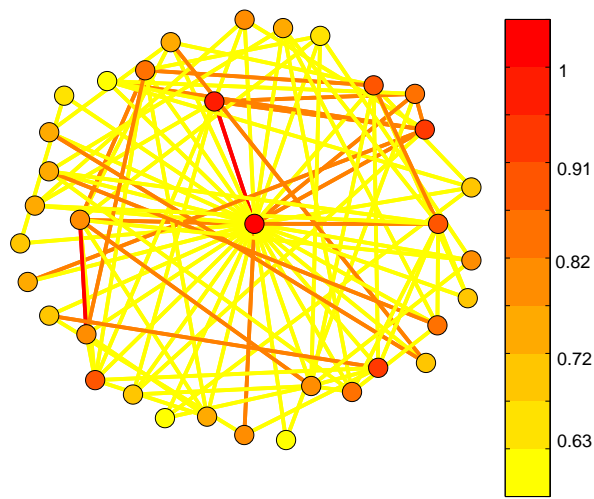}}
    \centerline{\parbox{320pt}{ \caption{
CMB net 1. Color represents the
probability $P^{+}$ to make decision +1. $P^{+}$ ranges in [0 1]. $P^{+}$
is computed from 100 runs. Dark edges represent a board interlock
consisting of two or more shared directors.
Clear edges represent one shared director interlock.
     }\label{ChaseNet1SpreadColor_a2_01}
     }}
\end{figure}

\textbf{Figure \ref{ChaseNet1SpreadColor_a2_01}} shows the map of the
probability $P^{+}$ to make decision +1 over the topology of the network.
Dark nodes make decision +1 with high probability. Dark edges represent a
board interlock consisting of two or more shared directors, clear edges
represent one shared director interlock.

For CMB Net 2, there is no dependence of $P^{+}$ on $\alpha_{1}$ even for
$\alpha_{2}$ small. The dependence of $P^{+}$ on $\alpha_{2}$
(\textbf{fig. \ref{figP_ChNet2} }, circles) is as dramatic as for CMB Net
1. \newline

Of course the chosen schedule of meetings when information propagates
from the center to the periphery seems important. To test this, we
considered three different cases:
\begin{itemize}
  \item CMB's board meets first, then the
boards 1 degree of separation away from CMB's board meet with a fixed
schedule and finally the boards 2 degrees away meet with a fixed
schedule.
  \item CMB meets first and
then the other boards meet with a schedule randomly chosen at every run,
  \item the schedule of all boards, including the first, is randomly chosen
at every run
\end{itemize}

Results obtained in the three cases are shown in \textbf{figure
\ref{figP_ChNet2} }(circles, triangles and stars, respectively). The
dramatic effect of information observed for case 1 (circles) is
insensitive to the particular schedule chosen, inside layer 1 and inside
layer 2. In case 2 (triangles) the probability of making decision +1
remains small even when the influence of information is strong. In case 3
(stars) the probability of making decision +1 is basically not different
from the control value 0.5 occurring when adoption of decision +1 and -1
are equally probable.

\section{ Discussion and Conclusions }
We have presented two models for the decision making process of a network
of interlocked boards about a common topic with a binary choice $\pm 1$.
The opinions of directors of the boards are initially evenly distributed,
but the most central board meets first and takes the decision +1 by
convention.
\begin{itemize}
\item
Model 1 only assumes that agents are imitative and that directors with
multiple appointments arrive at a board meeting with an initial opinion
which favors the decision made in the last board meeting they
participated at.
\item
Model 2 further takes into account the enhanced mutual influence of
directors sharing appointments in outside boards and the decisions made
by interlocked boards.
\end{itemize}

Results of simulations with model 1 show that the average probability
$P^{+}$ that a board adopts the decision of the center node increases
linearly with the average fraction $\gamma$ of directors shared by any
two boards until it saturates. But the value of $\gamma$ observed in real
board networks ($\gamma=0.1$) corresponds to only 60 per cent chances to
make decision +1, compared to 50 per cent chances in absence of
interlock. Therefore if one assumes that board interlock only introduces
a bias in the initial opinions of interlocking directors, then the effect
of the interlock on the decision making dynamics is very small.

For Model 2, we find that common appointments in external boards
($\alpha_{1}>0$) affects very weakly the spreading dynamics. By contrast,
information about interlocked boards' decisions ($\alpha_{2}>0$) has a
dramatic impact on the average probability $P^{+}$ that a board adopts
the decision of the center. If we assume that external information has an
influence comparable to the influence of colleagues' average opinion
($\alpha2 \simeq 0.1$), we then find that almost certainly all boards end
up adopting the decision of the center ($P_{+}=.9$ for CMB Net 1,
$P_{+}=.9$ for CMB Net 2).

The order of boards meeting is of crucial importance: a meeting schedule
respecting the distance of the boards from the central board is a
necessary condition to have a large majority of boards making the same
decision. In fact, innovations in real board networks has been observed
\cite{Davis96} to start from a peripheral node and almost never from the
most central node. But the diffusion takes place only when the central
node adopts the innovation and it is then imitated by connected boards.
In our model the radiation from the center hypothesis is based on the
idea that the topic is discussed in a board when it is proposed by an
interlocked director who has been involved in the same discussion on
another board. Even if in real boards the meetings are not scheduled as
we have assumed in our model, one can argue that directors of central
boards usually know \textit{before} the meeting what their board is
likely to decide and through interlock this information can be taken into
account in the meetings of more peripheral boards even if they are
scheduled before the meeting of the central boards.

The models presented here are very stylized with respect to the real
mechanisms involved in a decision making process in a network of boards.
Our hypothesis about agents fall in the framework of the "herd behavior".
We have chosen to make the smallest number of hypothesis on the agents in
order to keep the behavior of the system completely comprehensible.

As a general conclusion we find that imitation of colleagues and opinion
bias due to the interlock are not sufficient to trigger an avalanche of
identical decisions over the network, whereas information about
interlocked boards' decisions is. But there is no need to invoke public
information about what \textit{all} boards have decided, nor any external
dramatic events.\newline

This model thus provides a simple endogenous mechanism to explain
the fact that boards of the largest corporations of a country can,
in the span of a few months, take the same decisions about general
topics such as investments and advertisements.

\section{Acknowledgments}
We thank Gerald Davis (Univ. of Michigan), for having kindly provided the
data of the US Fortune 1000 and for helpful comments. We also thank
Jacques Lesourne and Andrea Bonaccorsi for fruitful discussion.\newline
This work is supported by FET-IST department of the European Community,
Grant IST-2001-33555 COSIN.

\section{ Figures }

\end{document}